\documentclass[prd,twocolumn,groupedaddress,showpacs,nofootinbib]{revtex4}
\usepackage{graphicx}
\usepackage{dcolumn}% Align table columns on decimal point
\usepackage{bm}% bold math
\usepackage{amssymb}
\usepackage{mathrsfs}
\usepackage{amsmath}
\usepackage{epsfig}
\usepackage[dvips]{color}
\usepackage{hhline}

\begin{document}

\title{Pseudo-Majoron as Dark Matter}

\author{Pei-Hong Gu}
\email{peihong.gu@mpi-hd.mpg.de}

\author{Ernest Ma}
\email{ma@phyun8.ucr.edu}

\author{Utpal Sarkar}
\email{utpal@prl.res.in}

\affiliation{Max-Planck-Institut f\"{u}r Kernphysik, Saupfercheckweg
1, 69117 Heidelberg, Germany\\
Department of Physics and Astronomy, University of California,
Riverside, California
92521, USA\\
Physical Research Laboratory, Ahmedabad 380009, India}

\begin{abstract}

We consider the singlet Majoron model with softly broken lepton
number. This model contains three right-handed neutrinos and a
singlet scalar besides the standard model fields. The real part of
the singlet scalar develops a vacuum expectation value to generate
the lepton number violation for seesaw and leptogenesis. The
imaginary part of the singlet scalar becomes a massive
pseudo-Majoron to be a dark matter candidate with testability by
colliders, direct detection experiments and neutrino observations.

\end{abstract}


\maketitle

\section{Introduction}

The existence of non-baryonic dark matter \cite{dunkley2008}
indicates the necessity of supplementing the  $SU(3)_{c}^{}\times
SU(2)_{L}^{}\times U(1)_{Y}^{}$ Standard Model (SM) with new
ingredients. There have been many interesting proposals for the dark
matter. The simplest dark matter candidate seems to be a scalar
phantom \cite{sz1985,bpv2000,hllt2009}. It is a stable SM-singlet
and has a quartic coupling with the SM Higgs doublet. Through its
annihilation into the SM particles, the dark matter scalar can
contribute a desired relic density to our universe. The coupling of
the dark matter scalar to the SM Higgs also opens a window to verify
the dark matter property by colliders and direct detection
experiments.

Observations on solar, atmospheric, reactor and accelerator
neutrinos have established the phenomenon of neutrino oscillations
\cite{stv2008}. The required massive and mixing neutrinos also
implies the need for new physics beyond the SM, where the neutrinos
are massless. Currently the seesaw \cite{minkowski1977} mechanism
and its variants \cite{mw1980,rw1983} are the most popular scheme
for generating the small neutrino masses. If the neutrinos are
Majorana particles, lepton number must be broken by the neutrino
masses. For example, in the canonical seesaw model
\cite{minkowski1977} the lepton number is explicitly broken by the
Majorana masses of right-handed neutrinos, which have Yukawa
couplings with the SM lepton and Higgs doublets. The seesaw models
also accommodate the famous leptogenesis \cite{fy1986} mechanism to
explain the matter-antimatter asymmetry in our universe. In this
scenario, a lepton asymmetry is first produced by the lepton number
violating interactions for generating the neutrino masses and then
is partially converted to a baryon asymmetry through sphaleron
\cite{krs1985}.

In order to understand the lepton number violation for the seesaw
and the leptogenesis, we can consider more fundamental theories,
where the lepton number violation is induced by spontaneous breaking
of a local or global symmetry. The simplest proposal is to consider
the singlet Majoron model \cite{cmp1980}, where the global lepton
number will be spontaneously broken after a complex singlet scalar
develops a vacuum expectation value (VEV). The right-handed
neutrinos can obtain Majorana masses through their Yukawa couplings
with this singlet scalar. Associated with the spontaneous symmetry
breaking of the global lepton number, there is a massless Goldstone
-- Majoron. The Majoron could become a massive pseudo-Majoron in
some variants of the singlet Majoron model. For example, in the
presence of the Yukawa couplings of new colored fermions to the
singlet scalar, the Majoron could get a tiny mass through the color
anomaly \cite{kim1979,shin1987}. This Majoron thus becomes a
pseudo-Majoron, which provides a solution for the strong CP problem,
as the global lepton number symmetry now is identified with the
$U(1)_{\textrm{PQ}}^{}$ symmetry \cite{pq1977}. In this paper, we
extend the singlet Majoron model with softly broken lepton number.
We find the induced pseudo-Majoron can account for the dark matter.
Compared with the dark matter model of the scalar phantom
\cite{sz1985,bpv2000,hllt2009}, the pseudo-Majoron as the dark
matter will not provide new implications on the collider
phenomenology and the direct detection experiments. However, the
neutrino fluxes from the decay of the pseudo-Majoron will induce
distinguishable signals in the future neutrino experiments
\cite{cgit2009}.

\section{The model}

We extend the SM with two types of singlets: one is a scalar and the
other contains three right-handed neutrinos. As the SM leptons carry
a lepton number $L=1$, we assign $L=1$ for the right-handed
neutrinos and $L=-2$ for the singlet scalar. Within the
renormalizable context, we can write down the lepton number
conserving interactions involving the neutrinos,
\begin{eqnarray} \label{yukawa}
\mathcal{L}_{Y}^{}\supset - y_{\nu}^{}\bar{\psi}_{L}^{}\phi
N_{R}^{}-\frac{1}{2}h\xi\bar{N}_{R}^{c}N_{R}^{}+\textrm{H.c.}\,.
\end{eqnarray}
Here $\psi_L^{}$ and $\phi$, respectively, are the SM lepton and
Higgs doublets, $N_R^{}$ denotes the right-handed neutrinos and
$\xi$ is the singlet scalar. The lepton number conserving potential
of the scalar fields should be
\begin{eqnarray}
\label{potential}
V(\xi,\phi)&=&-\mu_{1}^{2}\xi^{\dagger}_{}\xi+\lambda_{1}^{}(\xi^{\dagger}_{}\xi)^{2}_{}-\mu_{2}^{2}\phi^{\dagger}_{}\phi
+\lambda_{2}^{}(\phi^{\dagger}_{}\phi)^{2}_{}\nonumber\\
&+&2\lambda_3^{}\xi^{\dagger}_{}\xi\phi^{\dagger}_{}\phi\,,
\end{eqnarray}
where $\lambda_{1,2}^{}>0$ and
$\lambda_3^{}>-\sqrt{\lambda_1^{}\lambda_2^{}}$ to guarantee the
potential bounded from below.

In the Yukawa interaction (\ref{yukawa}) and the scalar potential
(\ref{potential}), the lepton number is flexible to be either a
local or a global symmetry. In the present work, we are interested
in the global case. It is well known that the spontaneous symmetry
breaking of the global lepton number will induce a massless Majoron.
Alternatively, we take into account the soft breaking of the global
lepton number to make the massless Majoron become a massive
pseudo-Majoron. For simplicity, we only introduce the following soft
term,
\begin{eqnarray}
\label{potential2}
V_{soft}^{}&=&-\frac{1}{2}\mu_{3}^{2}(\xi^{2}_{}+\textrm{H.c.})\,.
\end{eqnarray}
Other soft terms such as the Majorana masses of the right-handed
neutrinos can be forbidden by appropriate discrete symmetries. For
example, we can impose a $Z_4^{}$ symmetry under which
\begin{eqnarray}
\phi\rightarrow \phi\,,\quad\quad \xi\rightarrow -\xi\,,\quad
f\rightarrow i f\,.
\end{eqnarray}
Here $f$ stands for the SM fermions and the right-handed neutrinos.

It is convenient to expand the singlet scalar by two real scalars,
\begin{eqnarray}
\xi=\frac{1}{\sqrt{2}}(\sigma +i \chi)\,,
\end{eqnarray}
and then rewrite the full potential (\ref{potential}) and
(\ref{potential2}) in a new form,
\begin{eqnarray}
V&=&-\frac{1}{2}(\mu_1^2+\mu_3^2)\sigma^2_{}+\frac{1}{4}\lambda_1^{}\sigma^4_{}-\frac{1}{2}(\mu_1^2-\mu_3^2)\chi^2_{}\nonumber\\
&&+\frac{1}{4}\lambda_1^{}\chi^4_{}-\mu_{2}^{2}\phi^{\dagger}_{}\phi
+\lambda_{2}^{}(\phi^{\dagger}_{}\phi)^{2}_{}+\frac{1}{2}\lambda_1^{}\sigma^2_{}\chi^2_{}\nonumber\\
&&+\lambda_3^{}\sigma^2_{}\phi^{\dagger}_{}\phi+\lambda_3^{}\chi^2_{}\phi^{\dagger}_{}\phi\,.
\end{eqnarray}

\section{Vacuum Expectation Values and Masses of Scalars}

In order to break the lepton number and the electroweak symmetry,
the singlet scalar $\xi$ and the SM Higgs doublet $\phi$ will
develop VEVs. For example, with the following parameter choice,
\begin{subequations}
\begin{eqnarray}
\frac{\lambda_2^{}(\mu_1^2+\mu_3^2)-\lambda_3^{}\mu_2^2}{\lambda_1^{}\lambda_2^{}-\lambda_3^2}&>&0\,,\\
\frac{\lambda_1^{}\mu_2^2-\lambda_3^{}(\mu_1^2+\mu_3^2)}{\lambda_1^{}\lambda_2^{}-\lambda_3^2}&>&0\,,\\
\frac{\lambda_1^{}\lambda_2^{}\mu_3^2-\lambda_3^{2}\mu_2^2}{\lambda_1^{}\lambda_2^{}-\lambda_3^2}&>&0\,,
\end{eqnarray}
\end{subequations}
we find two nonzero and one zero VEVs,
\begin{subequations}
\begin{eqnarray}
u&=&\sqrt{2}\langle\sigma\rangle =\sqrt{
\frac{\lambda_2^{}(\mu_1^2+\mu_3^2)-\lambda_3^{}\mu_2^2}{\lambda_1^{}\lambda_2^{}-\lambda_3^2}}\,,\\
w&=&\sqrt{2}\langle\chi\rangle =0\,,\\
v&=&\sqrt{2}\langle\phi\rangle =\sqrt{
\frac{\lambda_1^{}\mu_2^2-\lambda_3^{}(\mu_1^2+\mu_3^2)}{\lambda_1^{}\lambda_2^{}-\lambda_3^2}}\,.
\end{eqnarray}
\end{subequations}
In consequence, there are three massive scalars
\begin{eqnarray}
\label{mass} \mathcal{L}_{m}^{}\supset
-\frac{1}{2}(\Phi,H,\chi)\left[
\begin{array}{ccc}
2\lambda_1^{}u^2_{}&2\lambda_3^{}uv&0\\
[3mm] 2\lambda_3^{}uv&2\lambda_2^{}v^2_{}&0\\
[3mm] 0&0&2\mu_3^2 \end{array} \right]\left[
\begin{array}{c}
\Phi\\
[3mm] H\\
[3mm] \chi\end{array} \right]\,,
\end{eqnarray}
where the scalars $\Phi$ and $H$ are defined by
\begin{eqnarray}
\sigma=\frac{1}{\sqrt{2}}(u+\Phi)\quad\textrm{and}\quad
\phi=\frac{1}{\sqrt{2}}\left[
\begin{array}{c}
v+H\\
[3mm] 0 \end{array} \right]\,.
\end{eqnarray}
The mass term (\ref{mass}) clearly indicates that $\chi$ has no
mixing with $\Phi$ and $H$ because of its zero VEV. The VEV $v\simeq
246\,\textrm{GeV}$ has been determined for the electroweak symmetry
breaking. We will clarify later the VEV $u$ should be near the GUT
scale as the pseudo-Majoron $\chi$ is expected to be an attractive
dark matter candidate. For such a huge hierarchy between $u$ and
$v$, the mixing between $\Phi$ and $H$ is extremely small so that
$H$ can be identical to the SM Higgs boson. We thus simplify the
mass term (\ref{mass}) to be
\begin{eqnarray}
\mathcal{L}_{m}^{}\supset
-\frac{1}{2}m_{\Phi}^{2}\Phi^2_{}-\frac{1}{2}m_{H}^{2}H^2_{}-\frac{1}{2}m_\chi^2\chi^2_{}
\end{eqnarray}
with
\begin{eqnarray}
m_{\Phi}^{2}\simeq 2\lambda_1^{}u^2_{}\,,~~ m_{H}^{2}\simeq
2(\lambda_2^{}-\frac{\lambda_3^2}{\lambda_1^{}})v^2_{}\,,~~
m_\chi^2=2\mu_3^2\,.
\end{eqnarray}

\section{Leptogenesis and Seesaw}

Before the spontaneous symmetry breaking of the lepton number, the
right-handed neutrinos don't have Majorana masses. Instead, they
have Yukawa couplings with the singlet scalar as shown in Eq.
(\ref{yukawa}). After the singlet scalar develops its VEV, we can
obtain the Majorana masses of the right-handed neutrinos. At this
stage, we can conveniently derive
\begin{eqnarray} \label{yukawa2}
\mathcal{L}_{}^{}\supset - y_{\nu}^{}\bar{\psi}_{L}^{}\phi
N_{R}^{}-\frac{1}{2}M_N^{}\bar{N}_{R}^{c}N_{R}^{}+\textrm{H.c.}
\end{eqnarray}
with the definition
\begin{eqnarray} \label{mass2}
M_N^{}=\frac{1}{\sqrt{2}}hu\,.
\end{eqnarray}
Clearly, with the Yukawa and mass terms (\ref{yukawa2}), the decays
of the right-handed neutrinos can generate a lepton asymmetry stored
in the SM lepton doublets as long as CP is not conserved
\cite{fy1986,lpy1986}. Subsequently the sphaleron \cite{krs1985}
process will partially transfer the produced lepton asymmetry to a
baryon asymmetry which accounts for the matter-antimatter asymmetry
in the universe. We should keep in mind that through the Yukawa
interaction (\ref{yukawa}), the right-handed neutrinos not only
obtain Majorana masses $M_N^{}$ but also couple to the scalars
$\Phi$ and $\chi$. The couplings to $\Phi$ and $\chi$ will result in
some annihilations of the right-handed neutrinos \cite{gs2009}. For
a successful leptogenesis, these annihilations should go out of
equilibrium before the out-of-equilibrium decays of the right-handed
neutrinos. Fortunately, in the presence of a huge lepton number
breaking scale $u\sim 10^{16}_{}\,\textrm{GeV}$, the Yukawa coupling
$h$ is much smaller than unit for the desired Majorana masses
$M_N^{}<10^{14}_{}\,\textrm{GeV}$ so that the annihilations can be
highly suppressed.

The interactions (\ref{yukawa2}) also accommodates the seesaw
solution to the small neutrino masses. After the electroweak
symmetry breaking, the SM neutrinos $\nu_L^{}$ can obtain Majorana
masses,
\begin{eqnarray} \label{mass3}
\mathcal{L}_{m}^{}\supset
-\frac{1}{2}m_\nu^{}\bar{\nu}_L^{}\nu_L^c+\textrm{H.c.}\,.
\end{eqnarray}
The neutrino mass matrix $m_\nu^{}$ is given by the seesaw
\cite{minkowski1977} formula,
\begin{eqnarray}
m_\nu^{}=-m_D\frac{1}{M_N^{}}m_D^T\quad\textrm{with}\quad
m_D^{}=\frac{1}{\sqrt{2}}y_\nu^{}v\,,
\end{eqnarray}
where $m_D^{}$ is the Dirac mass matrix between the left- and
right-handed neutrinos. For $M_N^{}<10^{14}_{}\,\textrm{GeV}$, the
neutrino masses can be naturally small with $m_D^{}$ being at the
weak scale.

\section{Pseudo-Majoron as Dark Matter}

From the Yukawa interaction (\ref{yukawa}), we can derive the
coupling of the pseudo-Majoron to the right-handed neutrinos,
\begin{eqnarray}
\label{yukawa3} \mathcal{L}\supset
-\frac{i}{2\sqrt{2}}h\chi\bar{N}_R^{c}N_R^{}+\textrm{H.c.}\,.
\end{eqnarray}
Since $\chi$ is expected to be the dark matter, its lifetime should
be long enough. For this purpose, we consider the case of
$m_{\chi}\ll M_N^{}$, which can be easily achieved by taking proper
parameter choice. In this case, the pseudo-Majoron will decay into
the SM particles through the virtual right-handed neutrinos.
Conveniently, we can integrate out the heavy right-handed neutrinos
to derive the effective couplings of the pseudo-Majoron to the
left-handed neutrinos,
\begin{eqnarray}
\label{yukawa4} \mathcal{L}_{eff}^{}=
i\frac{m_\nu^{}}{2u}\chi\bar{\nu}_L^{}\nu_L^c\left(1+\frac{H}{v}\right)^2_{}+\textrm{H.c.}\,.
\end{eqnarray}
If the pseudo-Majoron is in the TeV region, its decay width will be
dominated by the two-body decay into the left-handed neutrinos,
\begin{eqnarray}
\Gamma_{\chi}^{} =\frac{1}{16 \pi} \frac{\textrm{Tr}(m_\nu^\dagger
m_\nu^{})}{u^{2}_{}}m_{\chi}^{}=\frac{1}{16 \pi}
\frac{\Sigma_i^{}\hat{m}_{\nu_i^{}}^2}{u^{2}_{}}m_{\chi}^{}\,,
\end{eqnarray}
which, in turn, gives the lifetime,
\begin{eqnarray}
\tau_{\chi}^{}=\frac{1}{\Gamma_{\chi}^{}}&=&\left(\frac{0.01\,\textrm{eV}^2_{}}{\Sigma_i^{}\hat{m}_{\nu_i^{}}^2}\right)\left(\frac{u}{
5.5\times10^{15}_{}\,\textrm{GeV}}\right)^2_{}\left(\frac{\textrm{1\,TeV}}{m_\chi^{}}\right)\nonumber\\
&&\times 10^{26}_{}\,\textrm{sec}\,.
\end{eqnarray}
It is straightforward to see the lifetime can be very long for a
huge lepton number breaking scale.

As a successful dark matter candidate, its relic density should be
consistent with the cosmological observations \cite{dunkley2008}. It
has been studied by many people \cite{sz1985,bpv2000,hllt2009} that
a stable SM-singlet scalar with a quartic coupling to the SM Higgs
doublet can serve as the dark matter because it contributes a
desired relic density through the annihilations into the SM
particles. In the present model, the pseudo-Majoron $\chi$ also has
a quaric coupling with the SM Higgs,
\begin{eqnarray}
V\supset\lambda_3^{}\chi^2_{}\phi^\dagger_{}\phi\Rightarrow
\lambda_3^{} v \chi^2_{} H+\frac{1}{2}\lambda_3^{}
\chi^2_{}H^2_{}\,.
\end{eqnarray}
This means the unstable pseudo-Majoron $\chi$ with a very long
lifetime definitely can play the role of the dark matter. Note for a
sizable $\lambda_3^{}$, the present symmetry breaking pattern, i.e.
$u\sim 10^{16}_{}\,\textrm{GeV}\gg v\simeq 246\,\textrm{GeV}$,
requires a large cancelation between $\lambda_3^{}u^2_{}$ and
$\mu_2^2$ so that $\lambda_3^{}u^2_{}-\mu_2^2$ can be of the order
of $v^2$.

The coupling of the pseudo-Majoron $\chi$ to the SM Higgs $H$ not
only determines the dark matter relic density but also opens a
window for the dark matter detection \cite{sz1985,bpv2000,hllt2009}.
For example, the Higgs $H$ can mostly decays into the dark matter
$\chi$ so that the dark matter is possible to find as a missing
energy at colliders such as the CERN LHC. Furthermore, through the
t-channel exchange of the Higgs $H$, the dark matter $\chi$ will
result in an elastic scattering on nucleons. Therefore its property
can be verified by the dark matter direct detection experiments.
Actually, such a type of scalar dark matter can well explain the
recent CDMS-II \cite{ahmed2009} discovery \cite{hllt2009}. On the
other hand, the pseudo-Majoron $\chi$ couples to the right-handed
neutrinos $N_R^{}$. This makes the dark matter $\chi$ unstable.
Although the dark matter decays are highly suppressed by the huge
lepton number breaking scale, it can produce sizable neutrino fluxes
which are sensitive in the next generation of neutrino experiments
\cite{cgit2009}.

\section{Summary}

In the singlet Majoron model, there would be a massless Majoron
associated with the spontaneous symmetry breaking of the global
lepton number. In this paper, we extend the singlet Majoron model
with the softly broken lepton number so that the massless Majoron
can become a massive pseudo-Majoron. We demonstrate that this
pseudo-Majoron can be a dark matter candidate with interesting
implications on the collider phenomenology, the dark matter direct
detection experiments and the future neutrino observations.

\vspace{5mm}

\textbf{Acknowledgement}:  PHG thanks Manfred Lindner for
hospitality at Max-Planck-Institut f\"{u}r Kernphysik and thanks the
Alexander von Humboldt Foundation for financial support.

\end{document}